\documentclass[aps,nofootinbib,prl,superscriptaddress,tightenlines,notitlepage,
twocolumn,showpacs,floatfix]{revtex4-1}
\usepackage[colorlinks]{hyperref}
\usepackage{tabularx}
\usepackage{bm}
\usepackage{graphicx}
\usepackage{amssymb,bm,tensor}
\usepackage{xcolor}
\usepackage{amsmath}
\usepackage[varg]{txfonts}
\usepackage{enumerate}
\usepackage[normalem]{ulem}
\usepackage{bm}
\usepackage[OT1]{fontenc}

\begin{document}

\title{Granular mass perturbations on the pulsar -- supermassive black hole system}

\date{\today}

\author{Zexin Hu}
\affiliation{Department of Astronomy, School of Physics, Peking University, 
Beijing 100871, China}
\affiliation{Kavli Institute for Astronomy and Astrophysics, Peking University, 
Beijing 100871, China}

\author{Lijing Shao}\email{lshao@pku.edu.cn}
\affiliation{Kavli Institute for Astronomy and Astrophysics, Peking University, 
Beijing 100871, China}
\affiliation{Max-Planck-Institut f\"ur Radioastronomie, Auf dem H\"ugel 69, D-53121 Bonn, Germany}
\affiliation{National Astronomical Observatories, Chinese Academy of Sciences, 
Beijing 100101, China}

\begin{abstract}

Discovery and timing observations of a radio pulsar orbiting around Sagittarius~A*, the supermassive black hole (SMBH) in our Galactic Centre (GC), will provide unprecedented opportunities of studying the SMBH spacetime, testing gravity theories, and probing the astrophysical environment in the GC. However, unknown mass distributions might cause timing residuals that are much larger than the timing precision. With extensive numerical simulations, for the first time we find that the perturbations caused by a granular cusp of stellar-mass black holes in the GC lead to post-fit timing residuals of 10--100\,s---contrary to traditional wisdom---even for a pulsar in a tight orbit with an orbital period $P_b=0.5\,{\rm yr}$. Such a large timing residual can lead to significant measurement bias or even prevent construction of a phase-connected timing solution for the full orbit. We revisit the idea of extracting SMBH parameters only with data around periastron where the perturbation is small. Under the realistic phase-disconnected assumption, we point out that it is vital to consider the frame-dragging effect in the light propagation, which breaks parameter degeneracy and leads to an order of magnitude improvement for the measurement precision of the SMBH spin.

\end{abstract}

\maketitle

\textit{Introduction}---As a key science goal of several next-generation radio telescopes, such as the Square Kilometre Array (SKA)~\cite{Kramer:2004hd, Weltman:2018zrl, Schoedel:2024} and the next-generation Very Large Array (ngVLA)~\cite{Bower:2018mta, Wilner:2024bbb}, the discovery and timing observations of a radio pulsar closely orbiting around Sagittarius~A* (Sgr~A*), the supermassive black hole (SMBH) dwelling at our Galactic Centre (GC), will provide unprecedented tests of general relativity (GR)~\cite{SKAOPulsarScienceWorkingGroup:2025syv, Shao:2014wja}. It also enables precision measurement of the SMBH~\cite{Liu:2011ae, Psaltis:2015uza, Zhang:2017qbb, DellaMonica:2023ydm, Hu:2024blq, Bambhaniya:2025qoe, Hu:2026zcb}, as well as its astrophysical environment such as the dark matter distribution in GC~\cite{Hu:2023ubk, Yu:2025apk, Shao:2025vmb}. The measurement of the SMBH spin is of particular interest as it is crucial for tests of the no-hair theorem and cosmic censorship conjecture~\cite{Kramer:2004hd, Liu:2011ae}. Previous studies suggest that timing observation of a radio pulsar with an orbital period $P_b\lesssim 0.5\,{\rm yr}$ and an orbital eccentricity $e\sim0.8$ could measure the spin and quadrupole moment of Sgr~A* to a relative precision of about $10^{-3}$--$10^{-2}$ over a 5-yr time span with an assumed timing precision of $1\,{\rm ms}$~\cite{Liu:2011ae,Psaltis:2015uza,Zhang:2017qbb,Hu:2023ubk}.

Despite theoretical models suggesting that there are a large population of pulsars  in the GC~\cite{Pfahl:2003tf,Zhang:2014kva,2020A&A...641A.102S}, current observations have only found seven whose projected distances are within $100\,{\rm pc}$ from Sgr~A*~\cite{Eatough:2013nva, Lower:2024sdi, Johnston:2006fx, Deneva:2009mx, Desvignes:2025hkk}. The lack of discovery in this region might be attributed to the strong scattering caused by the highly turbulent interstellar medium towards the GC. 
It is believed that future surveys at high frequency with better instrument sensitivity will reveal the missing pulsar population. 

For the proposed gravity tests, a pulsar in a very tight orbit 
around Sgr~A* is needed~\cite{Liu:2011ae}.
The strict requirement of a pulsar with such a tight orbit mainly comes from two aspects. One is that a tight orbit provides stronger relativistic effects, including the spin-orbit coupling effectively at the $1.5$\,post-Newtonian (PN) order and quadrupole at the 2\,PN order. The other consideration is that a wide pulsar orbit is affected by the complex astrophysical environment, such as the stellar cusp around Sgr~A*. As discussed by~\citet{Merritt:2009ex}, these perturbations might obscure the signal of spin-orbit coupling and quadrupole of the SMBH and make the desired tests of GR infeasible. In contrast, a tight orbit is expected to be dominated by the gravity of the SMBH and hopefully evades the complication from perturbations~\cite{Liu:2011ae, Psaltis:2015uza}. 

Studying the orbital motion of S-stars around Sgr~A* has provided a constraint on the extended mass distribution in the GC. At 1-$\sigma$ level, the  enclosed maximum mass (except the central SMBH) inside the S2 orbit is around $1000\,M_\odot$~\cite{GRAVITY:2024tth}. This mass is close to the predicted mass of the stellar cusp surrounding Sgr~A*, which, at the scale of S2 orbit, is mainly contributed by the stellar-mass black holes (BHs)~\cite{Zhang:2023cip}. Different from a smooth mass distribution like the dark matter, a granular one can cause much larger perturbations due to close encounters. It is predicted that continuous tracing of the S2 orbit will start to observe the granular mass perturbations if the extended mass is large enough and indeed mainly consists of stellar-mass BHs~\cite{Bordoni:2025mli}.

A possible way to evade the unknown perturbations for a pulsar with a moderate orbital period is to only use the timing data during the pulsar's periastron passages~\cite{Psaltis:2015uza}. For a pulsar in an eccentric orbit, the perturbations mainly affect the pulsar during the apocenter part of the orbit, where the pulsar spends its most time. In contrast, for the pericenter passage, the pulsar's orbit is dominated by the SMBH. However, the measurement of the SMBH spin largely relies on the secular precession caused by the spin-orbit coupling~\cite{Wex:1998wt}. Perturbations around the apocenter might prevent construction of a phase-connected timing solution for those periastron passages, leading to a worse measurement precision than expected~\cite{Liu:2011ae, Psaltis:2015uza, Zhang:2017qbb, Hu:2023ubk}.

In this {\it Letter}, we present the first study of the timing residuals caused by granular mass perturbations for pulsars in tight orbits around Sgr~A*. Compared to the expected timing precision of the SKA and ngVLA, we find surprisingly large post-fit timing residuals, which suggest large systematic biases compared to statistical uncertainties on parameters. It might even prevent the construction of a phase-connected timing solution for the full pulsar orbit. Based on the realistic phase-disconnected assumption, we revisit the idea of using periastron-only timing data to measure the SMBH spin. Our analytical and numerical results prove the importance of considering the frame-dragging (FD) effect in light propagation, which is not included in the timing models used in previous studies~\cite{Liu:2011ae, Psaltis:2015uza, Zhang:2017qbb, Hu:2023ubk}. Observation of this effect breaks a spin parameter degeneracy in periastron-only timing observation and enhances the spin precision by about an order of magnitude.

\textit{Cluster model and simulations}---As long predicted by stellar dynamics, there could exist a stellar cusp around Sgr~A* due to mass segregation and two-body relaxation~\cite{Peebles:1972}. In a general model, this cluster is composed of several different populations, including main-sequence stars, white dwarfs, neutron stars, and BHs~\cite{Alexander:2008tq}. We take a simplified model and only focus on the stellar-mass BH component as they provide the main granular mass perturbation~\cite{Bordoni:2025mli}. Low-mass stellar objects can be approximated by a smoother extended mass distribution, which mainly provides an additional periastron precession~\cite{Hu:2023ubk}. We assume the BH cusp to be composed of equal mass point particles with a power-law density distribution $\rho(r)\propto r^{-2}$ around the central SMBH~\cite{Alexander:2008tq}. For a thermalized cluster, this gives a distribution of the semi-major axis $a$ and eccentricity $e$ of the BH orbits as 
\begin{equation}
    n(a,e)=2eN/a_m e_m^2\,,\quad\quad a\in(0,a_m)\,,\ e\in(0,e_m) \,,
\end{equation}
where $a_m$ and $e_m$ are the upper cutoffs, and $N$ controls the total number of BHs. The orientations of the BH orbits are uniformly distributed. The number of BHs is adjusted to fulfill the current constraint from S2, namely that inside the apocenter of S2, $r_0=9.4\,{\rm mpc}$, the extended mass contributed by the BH cusp, $M_{\rm ext}$, should not significantly exceed $1000\,{M_\odot}$~\cite{GRAVITY:2024tth}. With the above distribution, as long as $a_m\geq r_0/(1-e_m)$, one then has $N=M_{\rm ext}a_m/m r_0$, where $m$ is the BH mass.

In our simulations, 
we choose $e_m=0.9$ and correspondingly, $a_m=10r_0$. Though BHs can have higher eccentricities as suggested by the eccentricity distribution of S-stars~\cite{Gillessen:2017jxc}, this cutoff does not alter our main results. In fact, a lower cutoff reduces BHs in highly eccentric orbits that can move across the pulsar orbit, and gives a relatively smaller perturbation. We study four cases with diffenerent $m$ and $M_{\rm ext}$ for the BH cusp, namely, $m=10\,{M_\odot}$ or $50\,M_\odot$ and $M_{\rm ext}=1000\,M_\odot$ or $100\,M_\odot$. The two cases with the larger $M_{\rm ext}$ roughly represent the upper limit of the granular mass perturbation, while the smaller $M_{\rm ext}$ stands for more optimistic cases.

We use Keplerian orbits for all BHs and only numerically integrate the pulsar's motion. The equation of motion for the pulsar reads
\begin{equation}
    \bm{a}_{\rm PSR}=\bm{a}_{\rm N}+\bm{a}_{\rm 1PN}+\bm{a}_{\rm 2PN}+\bm{a}_{\rm S}+\bm{a}_{\rm Q}+\bm{a}_{\rm P}\,,
\end{equation}
where $\bm{a}_{\rm N}$, $\bm{a}_{\rm 1PN}$, and $\bm{a}_{\rm 2PN}$ are the PN expansion terms to 2\,PN order, $\bm{a}_{\rm S}$ and $\bm{a}_{\rm Q}$ are the leading-order spin-orbit coupling and quadrupole terms of the SMBH, $\bm{a}_{\rm P}$ denotes the perturbation caused by the stellar-mass BHs where Newtonian gravity is assumed. Details of the timing model we developed are given in Ref.~\cite{Hu:2026zcb}.

\textit{Timing residuals}---Based on the numerically integrated orbital motion of the pulsar, we obtain realistic times of arrival (TOAs) from the pulsar-SMBH system in the presence of granular perturbations. For the purpose of this work, we use zero noise injection. Fitting the simulated TOAs with a timing model without perturbation will then give us the unabsorbed timing residuals caused by the perturbations. In real observations, a structured timing residual significantly larger than the timing precision usually suggests unmodeled physical effects. Ignoring them in the timing model leads to significant parameter estimation biases compared to statistical errors~\cite{Kramer:2021jcw}.

\begin{figure*}
  \centering
  \includegraphics[width=17cm]{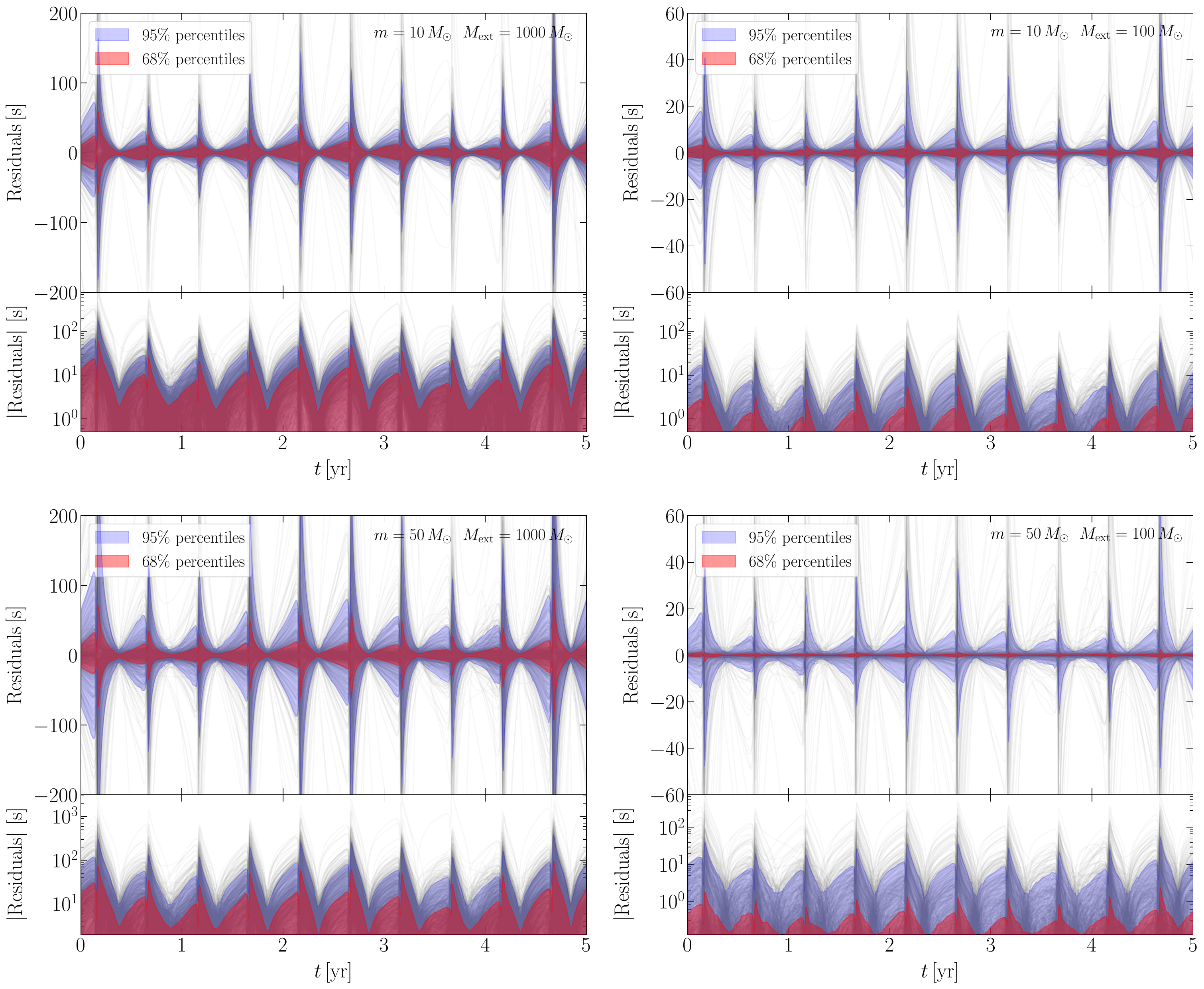}
  \caption{Post-fit timing residuals with different $m$ and $M_{\rm ext}$. Each gray line shows the result for one simulation, and contours  highlight the $68\%$ and $95\%$ percentile regions from 1000 simulations. We also show the absolute value of the residuals. The spike structures in the residuals are centered around periastron passages of the pulsar.
  \label{fig:residuals}}
\end{figure*}

To give a statistical point of view of the possible timing residuals caused by the perturbations from the BH cusp, we randomly generate $1000$ realizations of the BH cusp for each of the four cases mentioned before, while keeping same initial conditions for the pulsar orbit with $P_b=0.5\,{\rm yr}$ and $e=0.8$. We consider an observational time span of $5\,{\rm yr}$, which provides a precise measurement of the SMBH parameters when there is no perturbation~\cite{Liu:2011ae,Psaltis:2015uza,Zhang:2017qbb,Hu:2023ubk}. In Fig.~\ref{fig:residuals}, we show the post-fit timing residuals for all these simulations and highlight their $68\%$ and $95\%$ percentile regions. From the figure, one can see that for $M_{\rm ext}$ close to the upper limit given by the S2 star observation, the post-fit timing residuals can be as large as $10^2\,{\rm s}$. For $M_{\rm ext}=100\,M_\odot$, the residuals are smaller but still much larger than the expected timing precision, which for the SKA and ngVLA, despite the large scattering caused by the dense interstellar medium in the GC, is better than $\sigma_{\rm TOA}=1\,{\rm ms}$~\cite{Liu:2011ae}. 

We shall note that the post-fit timing residuals highly depend on the timing model that is used to fit the TOAs. Here we use the numerical timing model developed by \citet{Hu:2026zcb} that consistently includes all 2\,PN effects. We only fit for the pulsar's orbital and rotation parameters, as we will show that the SMBH parameters can be determined with periastron-only observations. Therefore, one can also regard the timing residuals in Fig.~\ref{fig:residuals} as under the condition where the SMBH parameters are unbiased. Additionally fitting  the SMBH parameters further reduces the residuals at the cost of introducing large biases in these parameters, which might be the case in real observations if the perturbations are not properly treated. Our results also suggest that, for an even tighter pulsar orbit like $P_b=0.1\,{\rm yr}$, the post-fit timing residuals are not significantly smaller.

In real timing observations, one does not know pulsar's true rotation phase at the time of each TOA. Therefore, the timing residuals in real observation will be folded into one pulsar rotation period~\cite{Damour:1986, Hobbs:2006cd}, which is ${\cal O}(1)\,{\rm s}$ for a normal pulsar and shorter for millisecond pulsars. The timing residuals shown in Fig.~\ref{fig:residuals} are phased-connected residuals, which means that we correctly add phase jumps to unwrap the folded timing residuals. However, considering the larger timing residuals compared to the pulsar's rotation period, finding a phase-connected solution is quite challenging in reality. 

\textit{Periastron-only analysis}---As shown, without further development of the timing model to include the granular mass perturbations, measuring the SMBH properties with data from the full pulsar orbit is problematic. \citet{Psaltis:2015uza} suggested using periastron-only data to constrain the SMBH parameters, which might alleviate the perturbations. Here we revisit this idea. With numerical simulations, we show that the granular mass perturbations  are indeed negligible in the periastron-only data. However, under a strong perturbation, we have shown that a phase-disconnected model should be adopted between orbits, giving a different expectation on the parameter measurement precision compared to Ref.~\cite{Psaltis:2015uza}. 

\begin{figure}
  \centering
  \includegraphics[width=8.5cm]{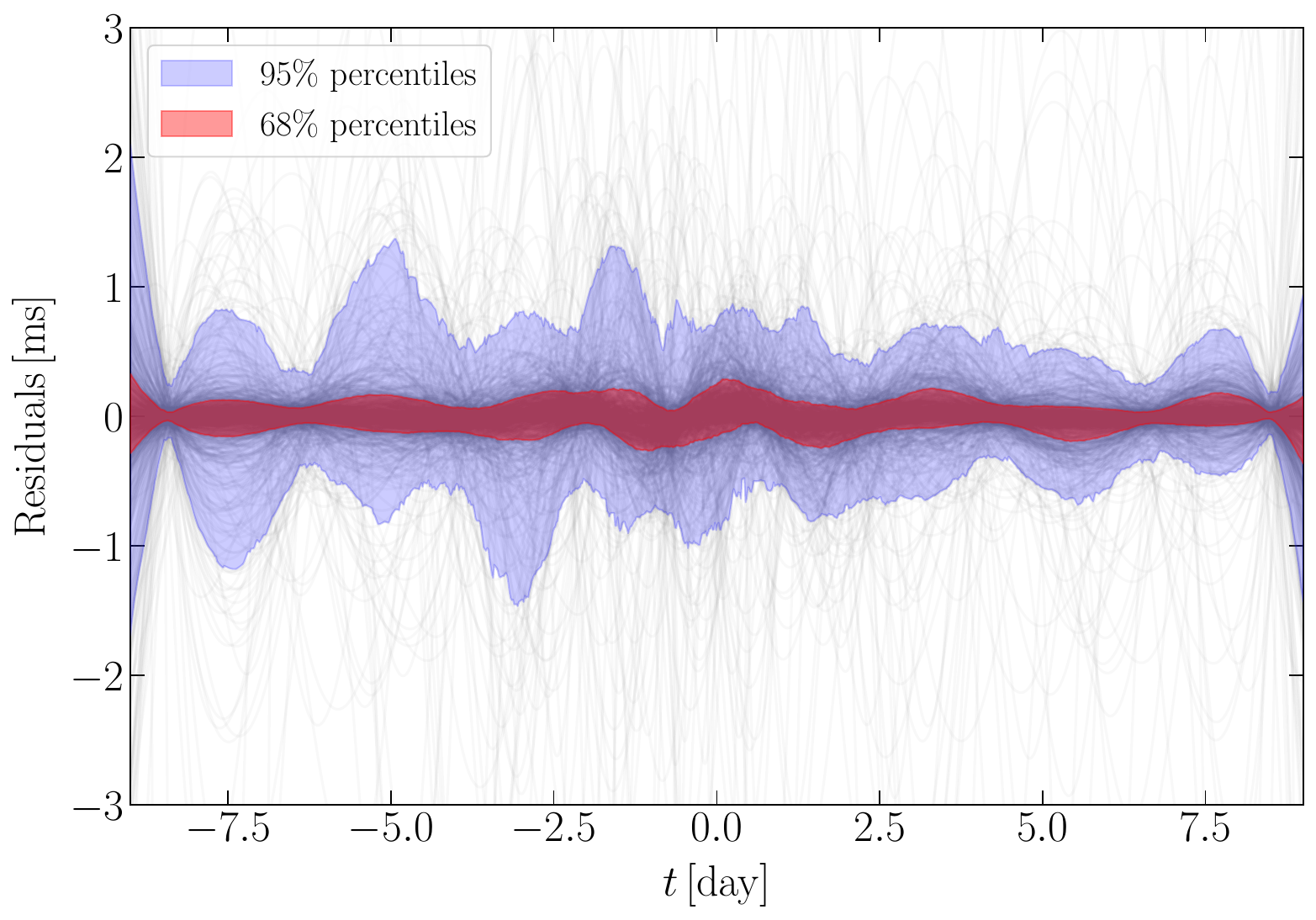}
  \caption{Post-fit timing residuals for periastron-only observation. We use a pulsar with $P_b=0.5\,{\rm yr}$ and $e=0.8$, and a BH cusp with $m=10\,M_\odot$ and $M_{\rm ext}=1000\,M_\odot$. The observation time window is $\pm0.05 \, P_b$ centered around the periastron passage. 
  \label{fig:peri_residual}}
\end{figure}

When studying the possibility of using periastron-only timing observations to measure the SMBH properties, we assume that the stellar perturbation is negligible, which is true for a pulsar in a highly eccentric and tight orbit. In Fig.~\ref{fig:peri_residual}, we show the post-fit timing residuals for the periastron-only timing observation of a pulsar with  $P_b=0.5\,{\rm yr}$ and  $e=0.8$ under the perturbation of a BH cusp with $m=10\,M_\odot$ and $M_{\rm ext}=1000\,M_\odot$. The observation time window is about $\pm 0.05\,P_b$ around the periastron as in Ref.~\cite{Psaltis:2015uza}. Our results suggest that the post-fit timing residuals for periastron-only data are less structured and below $\sim 1\,{\rm ms}$ for the case even when the full orbit suffers a large perturbation. This timing residual is comparable to the timing precision, and for a smaller $M_{\rm ext}$, the situation is even better. Therefore, the periastron-only observations indeed largely alleviate the perturbations. Note that, for this figure, we fit the full timing model including SMBH parameters as will be done in reality.

We have shown that it is hard to find, and also improper to use, a phase-connected timing solution to describe the pulsar's full orbit. A similar conclusion can be drawn for periastron-only timing data. Due to the perturbations during the apocenter part of the orbit, each time the pulsar moves back to its pericenter, the orbital parameters of the pulsar are changed. As suggested by the large timing residuals in Fig.~\ref{fig:residuals}, this change is much larger than the statistical uncertainties. Therefore, to combine the data of multiple periastron passages, one should regard orbital parameters independent for each segment. This argument simply leads to the fact that, the measurement precision of the SMBH parameters scale with the observed number of periastron passages as $N_{\rm peri}^{-1/2}$.

\begin{figure}
  \centering
  \includegraphics[width=8.5cm]{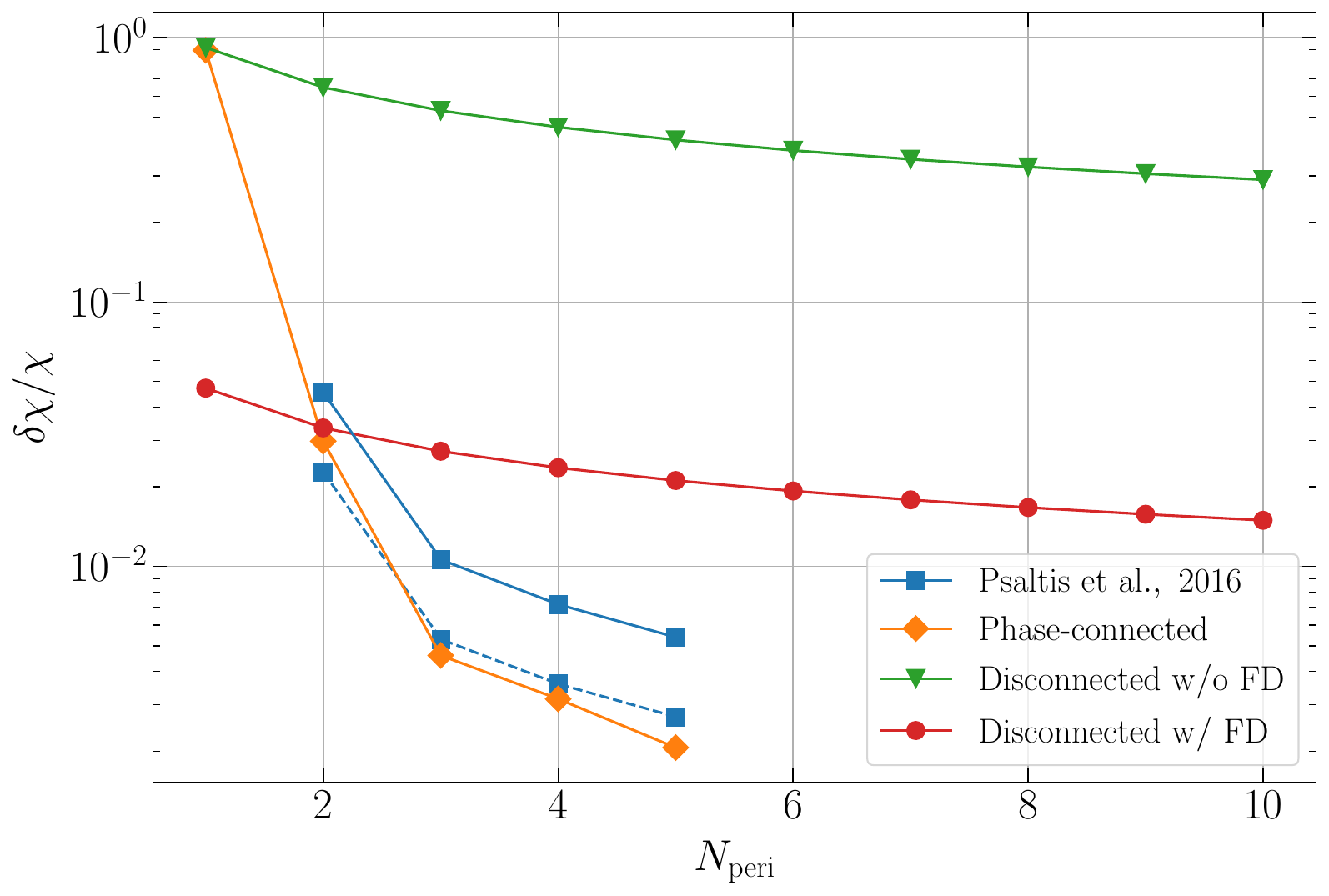}
  \caption{Fractional measurement precision of the dimensionless SMBH spin $\chi$ versus the numbers of periastron passage. The timing precision is assumed to be $\sigma_{\rm TOA}=100\,{\rm \mu s}$. We show 1-$\sigma$ results so that the solid blue curve is a factor of 2 smaller than the original curve in Fig.~7 of~\citet{Psaltis:2015uza}. The blue dashed line is another factor of 2 smaller  compared with the trend of our result. 
  \label{fig:spin}}
\end{figure}

We assume a timing precision of $\sigma_{\rm TOA}=100\,{\rm \mu s}$ as in Ref.~\cite{Psaltis:2015uza}. The observation is supposed to be dense so that we assume three TOAs per day. Based on the timing model and the Fisher matrix approximation developed~\cite{Hu:2026zcb}, we estimate the measurement precision of the SMBH parameters. In Fig.~\ref{fig:spin}, we show the fractional precision of the dimensionless spin $\chi$ of the SMBH. Along with the more realistic results for phase-disconnected analysis, we also show results based on the phase-connected assumption, that is, assuming the pulsar's orbital parameters unchanged among all periastron passages. One can see that our phase-connected result aligns well with the existing result~\cite{Psaltis:2015uza} except for an overall factor of two, which might come from the detailed difference in the timing model or parameter choice. The fast increase of the measurement precision across the first several periastron passages is a unique feature of (at least partially) phase-connected analysis, as it relies on the measurement of secular effects. The $N_{\rm peri}^{-1/2}$ scaling for the fully phase-disconnected results gives a much worse measurement precision as $N_{\rm peri}$ increases. Note that for a single periastron passage, the phase-connected and phase-disconnected models give same results.

Despite the phase-disconnected analysis giving a much worse measurement precision of $\chi$ than traditional expectation, we find that, including the FD effect in light propagation in the timing model, which was ignored in earlier studies, largely improves the result and leads to a spin measurement at the percent level as shown in Fig.~\ref{fig:spin}. This improvement, present in the numerical evidence, is possibly caused by the breaking of an approximate spin parameter degeneracy in the periastron-only observation. 

To understand this degeneracy, we may first look at a similar but strict degeneracy in the FD effect itself. The leading-order time delay of light propagation caused by the rotation of the SMBH is~\cite{Wex:1998wt}
\begin{equation}
    \Delta_{\rm FD}=-\chi\frac{2G^2M^2}{c^5}\frac{\hat{\bm s}\cdot\left(\hat{\bm K}_0\times\hat{\bm n}\right)}{r-z}=f(S_x,S_y)\,,
\end{equation}
where $M$ is the mass of the SMBH, $\hat{\bm{s}}$ is the direction of the SMBH spin $\bm{S}$, and $\hat{\bm n}={\bm r}/r$ points from SMBH to the pulsar with $r$ the radial distance; $\hat{\bm K}_0$ is the line of sight direction pointing from the Earth to Sgr~A* and gives the $z$-direction so that $z=\hat{\bm K}_0\cdot{\bm r}$. It is clear that, neglecting the proper motion of Sgr~A* during the short periastron passage---which changes $\hat{\bm K}_0$---the time delay caused by the FD effect only depends on the SMBH spin components in the $x$-$y$ plane. Therefore, if only $\Delta_{\rm FD}$ is considered, it is impossible to measure $S_z$. Equivalently, there will be an exact degeneracy in the spin parameters $\{\chi,\lambda,\eta\}$, with $\lambda$ and $\eta$ giving the direction of $\bm{S}$~\cite{Hu:2026zcb}.

Similarly, if one only considers the time delay caused by the spin-orbit coupling, there is an approximate degeneracy in periastron-only analysis. The main contribution of the additional time delay caused by the spin-orbit coupling comes from the change of the R\"{o}mer delay and is roughly
$\Delta_{\rm SO}\sim\hat{\bm K}_0\cdot \big( \int dt\int dt'\ {\bm a}_{\rm S} \big)$,
where
\begin{equation}
    {\bm a}_{\rm S}=\chi\frac{6G^2M^2}{c^3r^3}\left[\hat{\bm s}\cdot(\hat{\bm n}\times{\bm v})\hat{\bm n}+\dot{r}(\hat{\bm n}\times\hat{\bm s})-\frac{2}{3}({\bm v}\times\hat{\bm s})\right]\,,
\end{equation}
with ${\bm v}=\dot{\bm r}$~\cite{Barker:1975ae}. In the duration of a pulsar in an eccentric orbit passing its periastron, the change in the direction $\hat{\bm{n}}$ and $\hat{\bm{v}}$ will be much larger than the relative change in the velocity $\delta v/v$. Around the periaston, one  has a negligible radial velocity $\dot{r}$ compared to $v$. Further, around the pericenter, one has $\hat{\bm{n}}\times \hat{\bm{v}}\approx \hat{\bm{L}}$ with $\hat{\bm{L}}$ the direction of the orbital angular momentum. We shall note that, strictly speaking, these assumptions are only valid for a small true anomaly. Therefore, the following analysis gives an approximation that depends on both the orbital eccentricity and observation time window. Taking into account the above approximations, one can put the time delay in the following form
\begin{equation}
    \Delta_{\rm SO}\propto \int dt\int dt'\ \left[({\bm S}\cdot\hat{\bm L})\hat{\bm K}_0+2(\hat{\bm K}_0\cdot\hat{\bm L}){\bm S}\right]\cdot\hat{\bm n}\,.
\end{equation}
Note that, despite the fact that there can be a small orbital plane precession caused by the spin-orbit coupling, and changes in the orbital parameters caused by perturbations, $\hat{\bm n}$ is roughly changing only in the fixed plane of the pulsar orbit during the periastron passage. This suggests that $\Delta_{\rm SO}$ is nearly independent of the  
combination,
$\big[({\bm S}\cdot\hat{\bm L})\hat{\bm K}_0+2(\hat{\bm K}_0\cdot\hat{\bm L}){\bm S}\big]\cdot\hat{\bm L}$.
Similar to the simpler but strict case of $\Delta_{\rm FD}$ we discussed before, this  leads to a degeneracy among spin parameters. Nevertheless, as the specific combination is clearly different from the combination $f(S_x,S_y)$ for $\Delta_{\rm FD}$, combining the two effects largely breaks the degeneracies and leads to a much better spin measurement. Numerical result for a single periastron passage with the Fisher matrix suggests a clear decrease in the correlation among spin parameters. The absolute correlation $(|q_{\chi,\lambda}|,|q_{\chi,\eta}|,|q_{\lambda,\eta}|)$ changes from $(0.999,0.995,0.998)$ to $(0.928,0.954,0.973)$ after considering the FD effect.

We note that, the FD effect shows a similar feature to the so-called longitudinal deflection delay (or simply binding delay)~\cite{Doroshenko:1995rm,Wex:1998wt}, which is related to the pulsar's rotation. Depending on the orientation of the pulsar's rotation axis, the longitudinal deflection delay can have a similar amplitude as the FD effect and affect the spin measurement. If this is the case, combining scintillation observation to constrain the pulsar rotation axis will be helpful.

\textit{Discussions}---In this {\it Letter}, for the first time, with detailed numerical simulations we estimate effects of the perturbations caused by the BH cusp around Sgr~A* for the timing observation of a pulsar orbiting around Sgr~A*, and reach the following important conclusions.
First, with current constraints from S2 star observations, a reasonable BH cusp around Sgr~A* can  result in timing residuals that are much larger than the expected timing precision. With such large residuals, finding a phase-connected timing solution for the full pulsar orbit is challenging in reality. 
Second, revisiting the idea of using the periastron-only data, our results suggest that ignoring the perturbations during periastron passage is reasonable for most cases, but pulse phases are disconnected for subsequent orbits. Third, while the phase disconnection results in a worse  measurement precision of the SMBH spin, we find that including the FD effect in light propagation breaks a degeneracy of spin parameters in periastron-only analysis, and improves the measurement precision by about an order of magnitude, leading to a fractional uncertainty at the percent level. 

We point out the difficulties caused by the large timing residuals from granular mass perturbations. In principle, these timing residuals that are much larger than the timing precision are all {\it signals} rather than {\it noises}, though a full model accounting for all the objects in the cusp is unrealistic due to the extremely large parameter space. Nevertheless, it would be worth developing a more comprehensive, probably hierarchical, timing model to resolve these residuals in future observations, as these residuals eventually encode detailed information about the astrophysical environment around Sgr~A*. Tough modeling every single BH is hard, some coarse-graining properties of the BH cusp like the total mass inside the orbit or the power-law index of the distribution may be extracted from the timing residuals in a statistical way. Further, considering that the main contribution of the perturbations comes from several relatively close encounters, including one or several additional orbits in the timing model may significantly improve the fitting while keeping the number of parameters under control. An explicit example with an interrupting intermediate-mass black hole was investigated~\cite{Hu:2026xuv}. We leave a detailed systematic analysis for future study.

We thank Norbert Wex for helpful discussions.
This work was supported by the National Natural Science Foundation of China (124B2056, 12573042), the National SKA Program of China (2020SKA0120300), the
Beijing Natural Science Foundation (1242018), the
Max Planck Partner Group Program funded by the Max Planck Society, and the High-performance Computing
Platform of Peking University.

\bibliography{refs}

\end{document}